# SCoPE: Evaluating LLMs for Software Vulnerability Detection

José Gonçalves[0009-0004-1038-8384], Tiago Dias[0000-0002-1693-7872], Eva Maia[0000-0002-8075-531X] and Isabel Praça[0000-0002-2519-9859]

Research Group on Intelligent Engineering and Computing for Advanced Innovation and Development (GECAD), Porto School of Engineering (ISEP), 4200-072 Porto, Portugal
`{jpsgs,tiada,egm,icp}@isep.ipp.pt`

**Abstract.** In recent years, code security has become increasingly important, especially with the rise of interconnected technologies. Detecting vulnerabilities early in the software development process has demonstrated numerous benefits. Consequently, the scientific community started using machine learning for automated detection of source code vulnerabilities. This work explores and refines the CVEFixes dataset, which is commonly used to train models for code-related tasks, specifically the C/C++ subset. To this purpose, the Source Code Processing Engine (SCoPE), a framework composed of strategized techniques that can be used to reduce the size and normalize C/C++ functions is presented. The output generated by SCoPE was used to create a new version of CVEFixes. This refined dataset was then employed in a feature representation analysis to assess the effectiveness of the tool's code processing techniques, consisting of fine-tuning three pre-trained LLMs for software vulnerability detection. The results show that SCoPE successfully helped to identify 905 duplicates within the evaluated subset. The LLM results corroborate with the literature regarding their suitability for software vulnerability detection, with the best model achieving 53% F1-score.

## 1 Introduction

The rapid development and adherence to Internet of Things (IoT) technologies [1], has made modern software systems much more intelligent and interconnected. However, it has also raised several new security concerns, which are only aggravated by the fact that IoT technologies operate much closer to the user state than traditional devices [2]. The detection of software vulnerabilities in source code positively contributes to the quality of software and when detected in early development stages, it can improve user trust and help avoid financial and time losses [3]. Thus, in the IoT landscape, the detection of vulnerabilities in C/C++ code is increasingly relevant, since this language is frequently used in these devices due to their small binaries size and manual memory management.

Despite the well-defined security best practices associated with C/C++ Programming Language (PL), they are frequently ignored by developers which may compromise IoT devices [4]. To avoid the laborious process of manually reviewing code to find vulnerabilities, the scientific community has turned to more intelligent approaches leveraging advanced machine learning models, such as Large Language Models (LLMs) for Software Vulnerability Detection (SVD). Many research works have been able to achieve great results using synthetic data, but their performance on real-world code is still far from desired [5][6]. As such, in the code vulnerability detection field, researchers are putting more effort into using datasets containing code extracted from real-world projects to train the detection models [4][7][8]. One of these datasets is the popular CVEFixes [9], which contains vulnerable and non-vulnerable code from several PL. However recent research leveraging LLMs for SVD has identified several flaws in the dataset and concluded that LLMs are not ready for this task [10].

This work presents SCoPE, which implements a set of strategized techniques aimed at reducing vocabulary and code size of C/C++ code. The output of this tool will help in refining the faulty CVEFixes C/C++ subset entries by processing them and providing a new source code representation for fine-tuning LLMs. The resulting representation allowed for the detection of the flawed entries suggested by the literature, being created a refined version of CVEFixes without those faulty data. To evaluate the tool's techniques, three pretrained LLMs were fine-tuned for SVD using the refined version of CVEFixes and an analysis comparing the performance of the models fine-tuned before and after applying the tool's transformations to the code was conducted. The results show that the tool's processing techniques were not able to significantly improve the model's performance by simplifying the source code, which supports literature claims of LLMs not being ready for SVD [10]. The main contributions of our work include:

- Creation of SCoPE, a data processing framework for C/C++ code, that encompasses word inflection, string replacement, code normalization and cleanup to assess if those techniques can improve the model's results. Both the framework and the refined version of the dataset are publicly available on Github[1].
- An analysis of the C/C++ code extracted from CVEFixes dataset for source code vulnerability detection, including the proposal of a fixed version of this extraction in its latest version available.
- An analysis of the tool's techniques to improve feature representation, based on pretrained LLMs fine-tuned for SVD.

The rest of the paper is organized as follows. Section 2 describes related work concerning vulnerability detection on C/C++ source code using the CVEFixes dataset. Section 3 describes the experimental setup including dataset and models utilized. Section 4 presents SCoPE, the analysis conducted, and potential threats to validity. Finally, section 5 summarizes the work done, highlighting the main conclusions and future work.

---

[1] https://github.com/jp2425/SCoPE

## 2   Related Work

With the recent advances in machine learning, there is a renewed interest in using the most recent models to detect vulnerabilities in source code. However, these models' efficacy hinges on the data used in training. With this in mind, Bhandari Guru et. al. [9] curated a dataset named CVEFixes, comprising code extracted from real-world projects, enriched with pertinent metadata pertaining to vulnerabilities. This dataset encompasses code written in multiple PLs, facilitating the development of models capable of detecting vulnerabilities across various PLs using a single dataset.

Several studies used this dataset to harness recent Artificial Intelligence (AI) advancements, particularly those taking advantage of recent developments in LLM architectures. Michael Fu et. al. [11] created a tool which can be integrated in Integrated Development Environment (IDE) software, empowering developers to automatically identify vulnerabilities, assess their severity, and receive suggestions for remediation. To achieve this, the authors used multiple models, including Bidirectional Encoder Representations from Transformers (BERT) models, to detect the existence of vulnerabilities in the source code. The results were promising, with the proposed solution showing 65% of accuracy in the vulnerability detection task, and in a survey conducted among programmers who trialed the tool, the results indicated widespread agreement regarding its utility. Sindhwad et. al. [12] also employed LLMs in their solution, using them to resolve a vulnerability, while employing Convolutional Neural Network (CNN) and Long Short-Term Memory (LSTM) based models for vulnerability detection. After removing unnecessary formatting and comments from the code, the authors created and evaluated all models, achieving over 90% F1-score in the task of vulnerability detection. However, despite using CVEFixes to train the models used in vulnerability resolution, the authors relied on the Juliet dataset [13] to train the vulnerability detection models, which comprises synthetic data, a factor that could potentially result in suboptimal performance in real-world scenarios.

Other works also used this dataset for vulnerability detection and fixing. Zimin Chen et. al. [14] tackled the problem of employing a small vulnerability fix dataset by applying the concept of transfer learning. The authors hypothesized that bug-fixing and vulnerability-fixing tasks are related, leveraging this insight to initially train the models with extensive bug-fix data and subsequently fine-tune them with a smaller vulnerability-fix dataset. Despite exhibiting a low accuracy, the authors showed this approach is promising and has the potential to enhance results with limited data.

Recent studies show promising results, although Purba et al. [10] claimed that LLMs do not perform well in SVD. They reached this conclusion after assessing four LLMs using two datasets: the CVEFixes dataset, which contains raw source code, and other containing code gadgets. During their experiments they noted issues with the CVEFixes dataset, suggesting that the use of flawed data from CVEFixes might have affected their findings negatively. Considering the issue found in their work regarding CVEFixes and the claim about LLMs not being ready for SVD, this study will tackle the shortcomings presented in their work by refining the CVEFixes C/C++ subset and use it in a feature representation analysis for evaluating the performance of LLMs for SVD.

## 3      Experimental Setup

This section describes experimental setup details, including the CVEFixes dataset, the LLMs and hyperparameters utilized for the feature representation analysis. The model's training and evaluation was conducted in a laptop running Windows 11 operating system, equipped with a $12^{th}$ Gen I7-12650HX 2.1 GHz CPU, 64 Gb of RAM and an RTX A4500 GPU. The following sections describe relevant details of the experimental setup.

### 3.1     Data Analysis

This section outlines the methodology employed in the analysis of the CVEFixes dataset. This dataset constitutes a collection of real-world code sourced from diverse projects, spanning multiple PLs, and organized in a relational database format. The dataset was curated by employing a method centered around the processing of code differentials to monitor file and method modifications aimed at rectifying vulnerabilities, and the authors enriched the dataset with pertinent details concerning the patched vulnerabilities, such as associated CVE and CWE identifiers. Other supplementary information about the extracted code was included, encompassing PL specifications, code complexity metrics, and more [9].

For this work, version 1.0.7 of the CVEFixes dataset[2] was used, which is the most recent version available at the time of writing. Notably, this version incorporates an increased amount of data associated with C/C++ code compared to its predecessors. To use the code within this dataset, it was necessary to extract it from the database. This involved first creating indexes on columns containing identifiers to reduce processing time, as without this optimization, query execution times were too long. For our analysis, the code at function level was used along with their respective labels, so once the indexes were created, the query presented in Fig. 1 was executed to extract the relevant data from the dataset.

```sql
SELECT method_change_id, code, before_change
FROM method_change
INNER JOIN file_change ON method_change.file_change_id = file_change.file_change_id
WHERE programming_language = 'C' OR programming_language = 'C++';
```

**Fig. 1.** SQL query used to extract data from the dataset

Following the execution of the previous query, a total of 15649 entries related to C/C++ code were extracted successfully. Among these entries, 6515 were related to vulnerable code while 9134 were labeled as non-vulnerable. However, previous studies using this dataset have highlighted processing issues, which introduced code changes unrelated to vulnerability detection, raising concerns about data quality [10]. This work will analyze the issues in this dataset more in-depth, since the processing techniques applied can be useful to detect non-relevant entries, due to the normalization and generalization of the source code.

---

[2] https://zenodo.org/records/7029359

### 3.2   Large Language Models and Finetuning

The analysis leverages three LLMs of different architecture families: (i) CodeBERT from RoBERTa family, (ii) CodeGPT from GPT-2 family, (iii) and NatGen from Flan-T5 family. CodeBERT is comprised of 125M parameters, with embeddings of up to 512 tokens and is pretrained on CodeSearchNet dataset encompassing 2.3M functions across six PLs. CodeBERT excels in tasks such as natural language code search and code documentation generation [15] and has also been proven to be suitable for other tasks [16]. CodeGPT incorporates 124M parameters, with embeddings of up to 1024 tokens and was pretrained on two PLs extracted from the CodeSearchNet dataset to support code completion and text-to-code generation tasks [17]. However, it can be further fine-tuned for common downstream tasks, such as sequence classification. Lastly, NatGen from the Flan-t5 family, follows an encoder-decoder architecture with a total of 220M parameters, with embeddings of up to 512 tokens. It was pretrained mainly on CodeSearchNet with a few additional data on C and C# collected from active Github projects. NatGen excels across various downstream tasks, including code translation, text-to-code generation, and bug repair [18].

In the analysis conducted, each model was fine-tuned to perform binary sequence classification, of vulnerable C/C++ code. After running multiple iterations of hyperparameters, the best set was determined to be 10 training epochs with a batch size of 16, learning rate at $2 \times 10^{-5}$ with 200 warmup steps. The models were evaluated each 200 steps, to ensure they were being well trained. Moreover, due to computational constraints, a decision was made to implement Low-Rank Adaptation (LoRA) [19] in the LLMs training, which is a Parameter Efficient Fine Tuning (PEFT) technique that reduces the number of trainable parameters, significantly decreasing the training time.

## 4   SCoPE

To address the problem of reliably applying some well-known transformations used in other works [20][21][22], the SCoPE framework has been developed. It processes source code at function-level and utilizes the antlr4 framework [23] with a publicly available grammar [24] to provide a minified version of the code with a set of transformations applied. This tool was designed with flexibility in mind, allowing users to select the transformations they wish to apply to the code, namely: (i) replacing programmer-defined function and variable names, (ii) replacing strings with a generic token, (iii) normalizing whitespace, (iv) removing comments, and (v) tokenizing code. By applying techniques that can remove comments or normalize code formatting, these transformations can help programmers create minified versions of their code or prepare source code for use in AI model training. It also uses the antlr4 error recovery functionality to try to recover from errors during the code parsing, something especially useful since the used grammar doesn't support processing code with macros.

SCoPE can be easily integrated into other solutions, like REST APIs or scripts that process datasets concurrently, since it only requires the source code of a function and the definition of the transformations to be applied. After processing the source code provided, it will output the processed function in the format defined by the user: token or text-based representation.

The following sections describe two important topics of this tool's evaluation: (i) the preparation of a corrected version of the CVEFixes C/C++ subset using SCoPE, and (ii) a feature representation analysis of the tool's code processing techniques in the context of pretrained LLMs fine-tuned for SVD. Finally, the evaluation results are discussed, and potential threats to validity are identified.

### 4.1   Data Processing

To prepare the data for model training, SCoPE was employed to standardize the data. This involved using techniques such as shortening variable and function names, removing comments, and eliminating error-handling-related strings. Additionally, text-based representation was chosen primarily because LLM models have their tokenize techniques, and these options were selected to reduce code size, thus enabling more code integration into the LLM models, which have limited token processing capacity. Furthermore, all entries that encountered errors during processing were appropriately marked, facilitating their distinction from others in subsequent stages.

After applying these processing techniques to the dataset, the resulting output was used to create a dataset with the processed functions. Out of the 15649 entries extracted from CVEFixes, 3614 had errors during the processing phase, all related to the usage of macros in the code. The remaining 12035 entries were processed seamlessly without encountering any issues.

During the analysis of the entries which were processed with no errors, it was found that the issue pointed out by previous work regarding duplicated values is more complex than previously thought. Within the extracted subset, 905 instances of duplicated entries were detected, a revelation facilitated by our processing methods. These duplicates can be categorized into three groups: entries with identical code but differing comments, entries with identical content, and entries with identical code but different programmer-defined variable or function names.

Consider the example shown in Fig. 2, where two entries in the original dataset have identical code but different comments. In our methodology, comments are considered irrelevant to vulnerability detection. Therefore, if the comments are discarded during processing, the final code will be the same. It is important to note that some of these duplicate entries are also tagged with inconsistent labels in the original dataset, so the model would have two entries with the same code but different labels. Such inconsistencies could affect model's performance, since it would be trained on the same code with different labels. A similar issue was found, where multiple entries with the same content were labeled differently.

```
1  /* Determine the nubmer of DST elements.  */        1  /* Determine the number of DST elements.  */
2  cnt = DL_DST_COUNT (s,1);                           2  cnt = DL_DST_COUNT (s,1);
```

**Fig. 2.** Example of duplicated entries with differences in comments

Entries with identical code where programmers simply change function or variable names between commits have also been identified as a problem. During data analysis, this issue is not easily identified without generalizing programmer-defined names, as the code may vary between commits but retains the same semantic meaning. It becomes particularly relevant when datasets are constructed using differentials to find changes between vulnerable and fixed versions. Typically, if a function is altered in a commit patching a vulnerability, dataset authors presume that the modified functions contributed to the vulnerability's presence. However, if programmers only rename functions or variables to enhance clarity, irrelevant functions may appear in the dataset. For instance, consider two entries with identical code but differing programmer-defined function names, as illustrated in Fig. 3. Those two entries have the same code, with different function names, however they have opposite labels.

```
1 int pos() { return ptr - start; }          1 int length() { return ptr - start; }
```

**Fig. 3.** Example of duplicated entries with changes in the function name

After the identification of the duplicated entries, one example of each duplicated entry was retained in the new version of the dataset, prioritizing the non-vulnerable one when multiple entries contained the same content but different labels. After the removal of the duplicates, the dataset ended up with 11573 entries, where 4612 were vulnerable samples. For the feature representation analysis, two versions of the dataset were created: one containing the original code, and the other containing the processed code.

To ensure data imbalance won´t affect the model's performance [25], both datasets were balanced to a 50-50 ratio, reducing their size to 9224 entries. Afterwards, the data was split into train, test and validation sets following an 80-10-10 split with stratification to ensure class balance across the sets. Then, all three sets were tokenized, truncated, and padded to the maximum input length of the models.

### 4.2 Feature Representation Analysis

The feature representation analysis consisted in comparing the performance of the selected models. They were fine-tuned on the refined dataset, before and after the application of the framework's transformations to obtain the minified version of the source code. The goal is to evaluate the impact of the different feature representations on the models' performance. Table 1 summarizes the results obtained presenting for each model, the achieved accuracy, precision, recall and F1-score. The models trained before applying the transformations have suffix "-B" and the models trained after suffix "-A".

Table 1. Performance comparison between the LLM models trained on the fixed dataset before and after data processing.

| Model | Accuracy (%) | Precision (%) | Recall (%) | F1-Score (%) |
|---|---|---|---|---|
| CodeBERT-B | 52 | 52 | 52 | 52 |
| CodeBERT-A | 52 | 52 | 52 | 51 |
| CodeGPT-B | 49 | 49 | 49 | 49 |
| CodeGPT-A | 49 | 49 | 49 | 49 |
| NatGen-B | 53 | 53 | 53 | 53 |
| NatGen-A | 53 | 54 | 53 | 53 |

The results obtained suggest that NatGen achieved the best performance with 53% accuracy and F1-score and CodeGPT worst with 49% on both metrics. The analysis shows that the average number of tokens representing the functions was reduced from 374 to 300, which means that the functions size was significantly decreased. This is important to ensure that functions fit in the embeddings of each model. Nevertheless, the results show that even though the data processing allowed for a decreased average number of tokens and normalization of the source code, reducing the vocabulary size, the framework's transformations had no significant impact on the model's performance, which supports literature claims that LLMs are not ready for SVD [10] .

### 4.3   Threats to Validity

The threats to validity of the results obtained can essentially be found in the data utilized to train the models. Since the process of identifying erroneous entries is automated, it is possible that some unidentified faulty data points may still exist in the dataset. Regarding the comparative analysis, since the fixed dataset is comprised of very few data points, it is possible that its conclusion might not hold true with larger datasets, where there is greater variety of samples.

## 5   Conclusion and future work

This work presents SCoPE, a novel code processing framework capable of reducing C/C++ vocabulary and code size. To tackle literature claims regarding flawed data in the popular dataset CVEFixes and the inadequacy of LLMs for SVD, this tool was designed to apply multiple processing techniques on code. SCoPE is publicly available on GitHub, so other research works can leverage it for data processing and analysis.

The tool was utilized to process the CVEFixes C/C++ subset. By implementing code transformations, it allowed the identification and removal of erroneous entries from the dataset, resulting in a refined version of it that is now publicly available on the GitHub repository. The refined dataset and the tool's transformations were evaluated in a feature representation analysis, which consisted in comparing the performance of three LLMs fine-tuned for SVD, using the data representation before and after applying the processing techniques. The results corroborate with the literature, since there was no significant performance improvement.

As future work, it would be interesting to tackle the threats to validity of the work conducted. The expansion of the data processing framework for other PL would also be important and the validation of more datasets would be a significant future contribution levering the tool developed which is publicly available.

**Acknowledgements.** The present work has received funding from projects UIDB/00760/2020 and UIDP/00760/2020.